\documentclass[12pt]{iopart}

\usepackage[]{graphicx}
\usepackage{dcolumn}
\usepackage{bm}
\usepackage{color}
\usepackage{sistyle}

\begin{document}

\title[Simple FET amplifier \today]{Very simple FET amplifier with a voltage noise floor less than
1~{nV$/\sqrt{\mathrm{Hz}}$}}

\author{Luca Callegaro, Marco Pisani, Alessio Pollarolo}

\address{Istituto Nazionale di Ricerca Metrologica (INRIM)\\
Strada delle Cacce, 91 - 10135 Torino, Italy }

\ead{l.callegaro@inrim.it}

\begin{abstract}
A field-effect transistor (FET) amplifier for small voltage signals
is presented. Its design is elementary and the construction can be
afforded by anyone. Despite its simplicity, with a voltage noise
less than \SI{1}{nV / \sqrt{Hz}}, it outperforms commercially
available integrated FET amplifiers. 
The amplifier has a gain flatness better than \SI{1}{dB} over \SI{1}{MHz} bandwidth;
it can be employed as a front-end for signal analyzers or signal recovery systems.

\end{abstract}


\maketitle



\section{Introduction}

The experimentalists often need a preamplifier for small voltage
signals with an high input impedance and the lowest possible noise.

Amplifiers based on commercial field-effect transistors (FET)
integrated circuits (IC) are very simple to construct, and provide a
number of friendly properties (internal compensation, wide
bandwidth, wide supply voltage range, high power-supply rejection
ratio, etc). However, they have equivalent input voltage noise not
lower than $\approx 5$ nV$/\sqrt{\mathrm{Hz}}$. On the other end,
the construction and trimming of amplifiers based on discrete FETs
(see e.g. Refs. \cite{Klein1979, Jefferts1989, Neri1991,
Howard1999, Ferrari2002, Zhang2006}) could be sometimes tough for
the general experimentalist.

In the following, we describe a very simple FET amplifier, which
construction simplicity is comparable to those of IC-based
amplifiers. It has a number of friendly properties and its voltage
noise floor is better than 1 {nV$/\sqrt{\mathrm{Hz}}$}.

\section{The circuit}
\begin{figure}[!ht]
    \begin{center}
    \includegraphics[width=3.5 in]{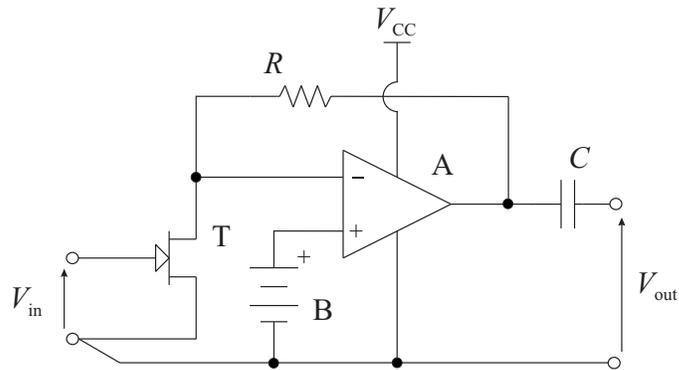}
    \caption{The schematic of the proposed amplifier. See text for an explanation of the symbols.}
    \label{fig:schema}
    \end{center}
\end{figure}

The circuit is shown in Fig.~\ref{fig:schema}.

The transistor T, directly connected to the input voltage
$V_\mathrm{in}$, is an FET in an common-source configuration. T
works with a gate-source bias voltage $V_\mathrm{GS} \approx 0$.
\footnote{A drawback of such configuration is the limited dynamic
range, a few tens of \SI{}{mV} at the input, before distortion
occurs.}

The wiring to $V_\mathrm{in}$ proposed in Fig.~\ref{fig:schema}
allows a a pseudo-differential configuration. This can help in
reducing interferences or, in correlation measurements
\cite{Callegaro2008}, permits a four-terminal connection to the
device under test.

The op amp A works as a transresistance amplifier with gain $R$, and
sets the FET transistor in a cascode configuration
\cite{Hunt39,Horowitz} (which eliminates the Miller effect
\cite{Miller20}, thus enhancing the bandwidth). The drain-source
voltage $V_\mathrm{DS}$ is set by A at the voltage $V_\mathrm{B}$
of\ the polarization battery B. B is practically unloaded
($I_\mathrm{B}$ is the small bias current of A), hence it has a long
life, and $V_\mathrm{B}$ has an extremely low noise
\cite{Boggs1995}.

A works in a single-supply configuration; its output $V_\text{out}$
is ac-coupled through capacitor $C$, and can be further amplified by
additional stages if necessary.

The overall low-frequency gain $G = + g_\mathrm{m}\,R$ of the
amplifier depends on $R$ and the FET transconductance
$g_\mathrm{m}$, which can have significant deviations from one
sample to another. Therefore, depending on the application, a
calibration of $G$ with a reference signal may become necessary
\cite{Callegaro2008}.

Several other properties of the amplfier, like the compensation,
bandwidth flatness, large supply voltage range, high power-supply
rejection ratio (PSRR), are given by A.

Typical supply voltage is $\mathrm{V_{CC}} = +24$ V from an
unregulated battery; the power load is $\approx$ \SI{15}{mA}.

\section{Experimental}
\label{sec:Experimental}

The circuit is simple to construct and does not require any
trimming.

Examples of suitable low-noise FETs are 2SK170 (Toshiba), LSK170
(Linear Systems), LSK389 (dual FET, Linear Systems);
Tab.~\ref{tab:datasheet} shows an extract of the corresponding
datasheets. A prototype has been assembled with a 2SK170, using an
OP27 (various suppliers) for A. We set $R=$\SI{1}{k\ohm}. The bias
current is $2\div 3$ \SI{}{pA} (measured with a Keithley mod. 6430
current meter), which gives a current shot noise less than $1$
\SI{}{fA/\sqrt{Hz}}.

\begin{figure}[!ht]
    \begin{center}
    \includegraphics[width=3.5 in]{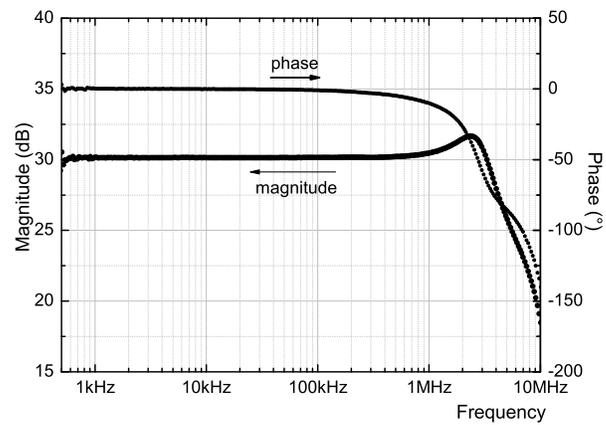}
    \caption{The transfer function of the amplifier, in magnitude and phase representation.}
    \label{fig:transfunc}
    \end{center}
\end{figure}

The transfer function of the amplfier is shown in
Fig.~\ref{fig:transfunc}. It has been measured with a network
analyzer (Agilent Tech.\ mod.\ 4395A), injecting the signal with a
resistive divider (\SI{50}{\ohm}-\SI{0.5}{\ohm}). A \SI{3}{dB}-bandwidth of
$\approx$\SI{4}{MHz} can be estimated, with a gain flatness better than \SI{1}{dB} up to 1 MHz.

\begin{figure}[!ht]
    \begin{center}
    \includegraphics[width=3.5 in]{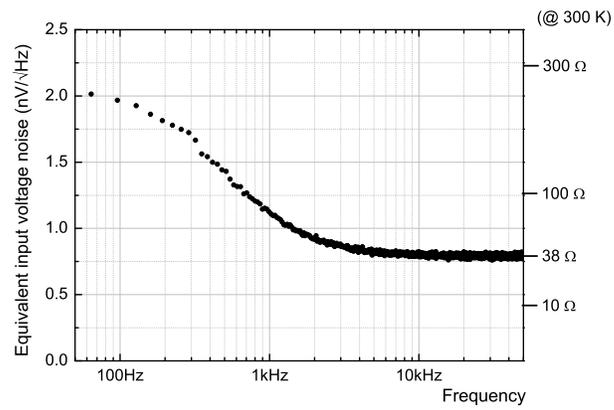}
    \caption{Equivalent input voltage noise of the amplifier.}
    \label{fig:noisefloor}
    \end{center}
\end{figure}

The equivalent input voltage noise with short-circuited input is
shown in Fig.~\ref{fig:noisefloor}. It has been measured with a
two-channel signal analyzer (Agilent Tech.\ mod.\ 35670A) by
connecting $V_\text{out}$ to both channels and performing a
cross-correlation measurement in order to reject the analyzer noise.
The noise floor is $\approx$\SI{0.8}{nV/\sqrt{Hz}}, corresponding to
the Johnson noise of a $\approx$\SI{38}{\ohm} resistor at
\SI{300}{K}.

\section{Conclusions}
The amplifier can be of interest in a number of applications, in
particular as an input front-end for commercial instrumentation
(signal analyzers, lock-in amplifiers).

Two protypes of the amplifier have been employed in a correlation
spectrum analyzer, developed for accurate measurement of the Johnson
noise of a \SI{1}{k\ohm} resistor \cite{Callegaro2008}. The
experiment is still under development, but the amplifier has shown
excellent properties, comparable with more sophisticated setups
\cite{Zhang2006}. Despite the open-loop configuration, after an
initial warm-up time the gain mid-term stability is within one part
in $10^3$ in a laboratory environment.

\section*{Acknowledgments}

The authors warmly thank Massimo Ortolano,
Politecnico di Torino, for fruitful discussions and for reviewing
the manuscript.

\newpage

\bibliography{FETAmpli}
\bibliographystyle{unsrt}

\newpage
\begin{table}[h]
    \caption{Main specifications of FET models suggested in Sec. \ref{sec:Experimental}, from manufacturer datasheets in typical operating conditions.}
    \centering
    \label{tab:datasheet}
    \begin{tabular}{c|c|c|c|c}
    \hline
        Symbol & description & 2SK170BL & LSK170B & LSK389B\\
    \hline\hline
    $g_\mathrm{m}$   &  transconductance & \SI{22}{mS}  & \SI{22}{mS} & \SI{20}{mS}    \\
    $e_\mathrm{n}$   &  noise voltage (\SI{1}{kHz}) &   \SI{0.95}{nV/\sqrt{Hz}}  & \SI{0.9}{nV/\sqrt{Hz}}  & \SI{0.9}{nV/\sqrt{Hz}}  \\
    $C_\mathrm{in}$ &  input capacitance    &   \SI{30}{pF} & \SI{20}{pF} & \SI{25}{pF}  \\
    $I_\mathrm{DSS}$ &  drain current    &   \SI{6\div12}{mA} & \SI{6\div12}{mA} & \SI{6\div12}{mA}  \\
    \hline
    \end{tabular}
\end{table}

\newpage

\end{document}